\documentclass[a4paper,11pt]{article}
\usepackage{graphicx}
\usepackage{dcolumn}
\begin{document}
\title{Equilibrium spherically curved 2D Lennard-Jones systems}
\author{J.M. Voogd,\footnote{Now at TNO-FEL, NL-2597AK The Haque, The Netherlands} 
\ P.M.A. Sloot\\ 
\small{University of Amsterdam, Section Computational Science,} \\ 
\small{Kruislaan 403, 1098 SJ, Amsterdam, The Netherlands}\\ 
\small{and}\\
\setcounter{footnote}{6}
R. van Dantzig\,\footnote{rvd@nikhef.nl, corresponding author} \\
\small{NIKHEF, Kruislaan 409, 1098 SJ Amsterdam, The Netherlands}}
\maketitle
\vspace{-5mm}
\begin{center} \large \bf \large Abstract\end{center}
\begin{center}
\parbox{12cm}{
\textit{To learn about basic aspects of nano-scale spherical molecular shells during their formation, spherically curved two-dimensional $N$-particle Lennard-Jones systems are simulated, studying curvature evolution paths at zero-temperature. For many $N$-values ($N<800$) equilibrium configurations are traced as a function of the curvature radius $R$. Sharp jumps for tiny changes in $R$ between trajectories with major differences in topological structure correspond to avalanche-like transitions. For a typical case, $N=25$, equilibrium configurations fall on smooth trajectories in state space which can be traced in the $E-R$ plane. The trajectories show-up with local energy minima, from which growth in $N$ at steady curvature can develop.}}
\end{center}
\section{Introduction}
{\em In vitro} self-organization in aqueous solution of nanoscale spherical shells, like various types of nano-vesicles\,\cite{biomembrane,liposomes,fullerene-plus,nanowheels}) 
and viral capsids\,\cite{capsids} is a thermodynamic process driven by overall free energy minimization. The resulting shell has inherent tendencies toward a) crystalline structure\,\cite{voogd-thesis,bowick,leech}, b) global polyhedral symmetry\,\cite{rapaport,icosa,northby}, and c) discrete
sizes\,\cite{biomembrane,marzec,reddy}, thus discrete curvatures.

A compelling question is whether during thermodynamic growth a transient uncompleted (open) shell of given $N$ prefers intrinsically to adopt a specific equilibrium radius of curvature ($R$), and whether this radius - on the basis of molecular packing order already - may be approximately stable during (part of) that growth. The occurrence of local minima in the mean (per particle) potential energy ($E$) of an emerging shell as a function of all internal degrees of freedom - including $R$ - along the evolution path in state space, could thus be significant for steady curvature during growth in $N$.
 
In this paper we approach the problem in a much simplified model that allows systematic generic studies. The sphericity of the shell, in reality due to intrinsic 3D-properties of the optimally closely packed molecular subunits, is built-in as a global geometrical constraint. We perform computer experiments with freely relaxing - possibly open - spherically curved zero-thickness monolayers of identical molecules, studying structural properties in relation with $N$ and $R$. Our approach is most natural for monolayer nanowheel vesicles\,\cite{nanowheels}, but it is as well relevant for tiny bilayer vesicles, where the laterally most densely packed pseudo-crystalline inner lipid headgroup-sublayer can act as monolayer `backbone'.
 
As model we use two-dimensional (2D) Lennard-Jones (LJ) $N$-particle systems on a spherical surface with flexible radius at zero temperature\,\cite{voogd-thesis,wette}. The LJ potential $V_{ij} = {r_{ij}^{-12ff}} - 2 {r_{ij}^{-6}}$ between two particles $i$ and $j$ with Euclidean distance $r_{ij}$ energetically favors close regular packing with essentially unit distance between neighboring particles\,\cite{voogd-thesis}. The LJ-form in the constrained system acts as an effective interaction mimicking the real complex of interactions. It allows a comprehensive systematic exploration while keeping salient features of real systems.
The present work arises from a series of computational studies\,\cite{voogd-thesis} on 2D spherical crystallization in LJ-systems over a broad range of $N$, involving thermodynamic behavior and zero-temperature global energy minimization. The LJ-systems follow local-equilibrium paths in state space, realistically allowing for local minimum 'hang-ups' in evolving configurations.

When many LJ particles are randomly spread over a flat surface they aggregate into an approximately homogeneous configuration, a major fraction of the
particles being trapped inside the bulk (interior). Edge particles have higher energy than bulk particles, giving rise to edge tension, the 2D equivalent of surface tension. Minimizing edge energy, flat aggregates become approximately circular patches. Minimizing the overall potential energy the bulk becomes an essentially regular hexagonal lattice. Allowing for spherical curvature, the 2D system can further decrease the energy by reducing the edge length and by a rising attraction from LJ-tails of remote particles (the LJ-forces acting in 3D). The energy gain by curvature, however, balances against increasing strain energy of the bulk because of less favorable packing. 

Our question becomes whether -- thanks to the interplay of these $R$-dependent non-linear effects -- equilibrium radii $R_{eq}$ can be found where relaxation
after any small change in curvature raises $E$, and freely variable $R$ leads the system back to the same equilibrium radius. If such local energy minima in {\em open} configurations are thermodynamically significant, they can stabilize transient states along a path of growing $N$.
This is supported by a LJ-study\,\cite{voogdp157}, showing that {\em closed} global minimum energy $N$-particle configurations (covering the whole sphere) strongly correlate with specific {\em open} local minimum configurations of lower $N$.

\section{Methods} 
In our experiments, each time the radius $R$ is changed by a small step, the particle system is relaxed by minimizing -- at the new $R$ value -- the mean  energy, which implies reaching the nearest equilibrium configuration. The system is thus evolved in curvature while staying in equilibrium with changing $R$. For {\em relaxation} (equilibration of forces and energy minimization), a steepest descent (SD) algorithm\,\cite{steepest-descent} is applied while for {\em aggregation} a Metropolis Monte Carlo simulated annealing (SA)\,\cite{laarhoven} optimization is used. The latter method enhances the probability that the system ends up in a {\em global} energy minimum (GEM)\,\cite{gem} rather than in a local secondary energy minimum. The 2D-topology of a configuration is defined by the Voronoi nearest neighbors method\,\cite{voronoi}, giving each particle a coordination number ($CN$), which is 6 everywhere for a flat (hexagonal) GEM configuration. The value $CN-6$ is denoted as the disclination charge - short d-charge - of a particle. In a 2D topological structure any single built-in defect -- disclination or dislocation (tightly bound pair of disclinations with opposite d-charge) -- can be displaced but not removed, except by its annihilation as part of a set of converging complementary defects, or by moving it all the way to the edge. During the transformation of a flat GEM lattice to a closed shell a net total d-charge of -12 must be incorporated in the full Voronoi lattice. In addition, dislocations have the function of lowering the  strain energy in total, by distributing it locally more evenly\,\cite{voogd-thesis}.      

How much the configurations change during relaxation after a step $\Delta$R, can be expressed as the mean Euclidean distance, $\hat{r}$, traveled between the associated sets of coordinates
${\bf x}_{R}$ and ${\bf x}_{R + \Delta R}$ in configuration space (mean taken per particle and per percent change in curvature):\\ $\hat{r} = \frac{| \Delta R |}{R N} \sqrt{\sum_{i=1}^{N} ({\bf x}_{R,i}-{\bf x}_{R + \Delta R,i})^2}$.

\section{Computer experiments}
\noindent{\center \em From flat to spherical\\} 
In a first series of experiment we explore gross changes in topological and geometrical structure and the corresponding energy with monotonously decreasing $R$. For hundreds of runs with $N<800$ an initial, circularlike aggregate is prepared from a flat regular hexagonal lattice with unit spacing (GEM for infinite $N$ in flat 2D). In the experiments $R$ is decreased in 1\% steps, each time the configuration being
projected onto the new sphere and then relaxed using the SD method. The decrease in $R$ is continued until the system is compressed considerably. A system of particular $N$ follows a `standard' (for that $N$) evolution path through state space and through the $E - R$-plane. 

\noindent{\center \em Example, $N=25$\\}
Secondly, curved lattices unbiased by any initial configuration and path history, are simulated while starting in SA mode at high temperature ($T=10$) in a random configuration and then aggregate by cooling down the system in 5\,\% $\Delta T/T$ steps. SD is applied as final tuning. This study is done for a {\em typical `unmagic'} $N$ value, $N=25$. In 1300 runs radii are randomly chosen between $R=1.3$ and $R=2$.

The equilibrium points in $E-R$ space align over a range of $R$-values along distinct smooth trajectories: lines which correspond to continuous sets of topologically and geometrically closely similar equilibrium configurations.

\noindent{\center \em Tracing up and down in curvature\\} The central question of the current paper, whether trajectories of the system lining-up closely related configurations, can provide a stable system against freely variable $R$, is addressed in a third type of experiment. Starting from particular open $N=25$ configurations obtained in the
second experiment, trajectories are traced (with relaxation) step-by-step in $R$ in both directions, and the structure is investigated.
 
\section{Results and discussion} 
\begin{figure}[t]
\begin{center}
\includegraphics[width=48mm]{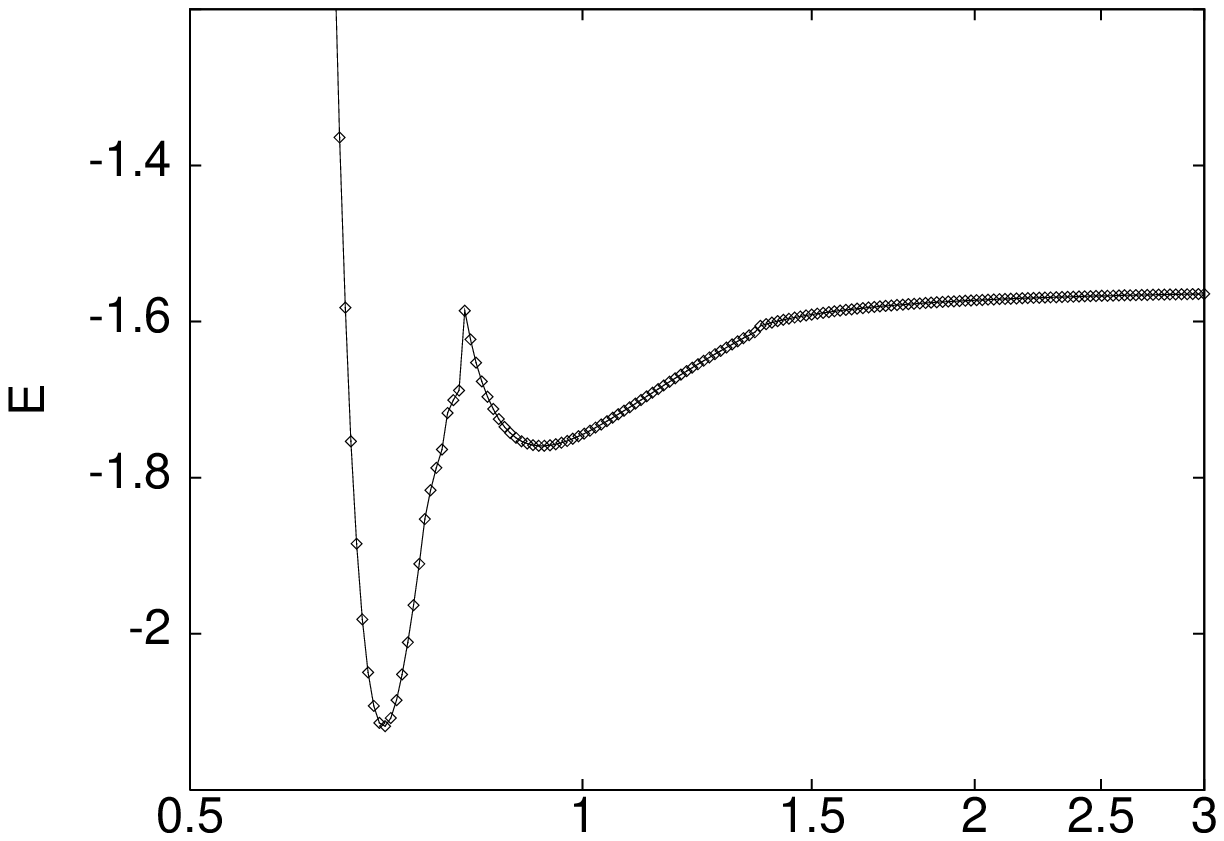}
\includegraphics[width=48mm]{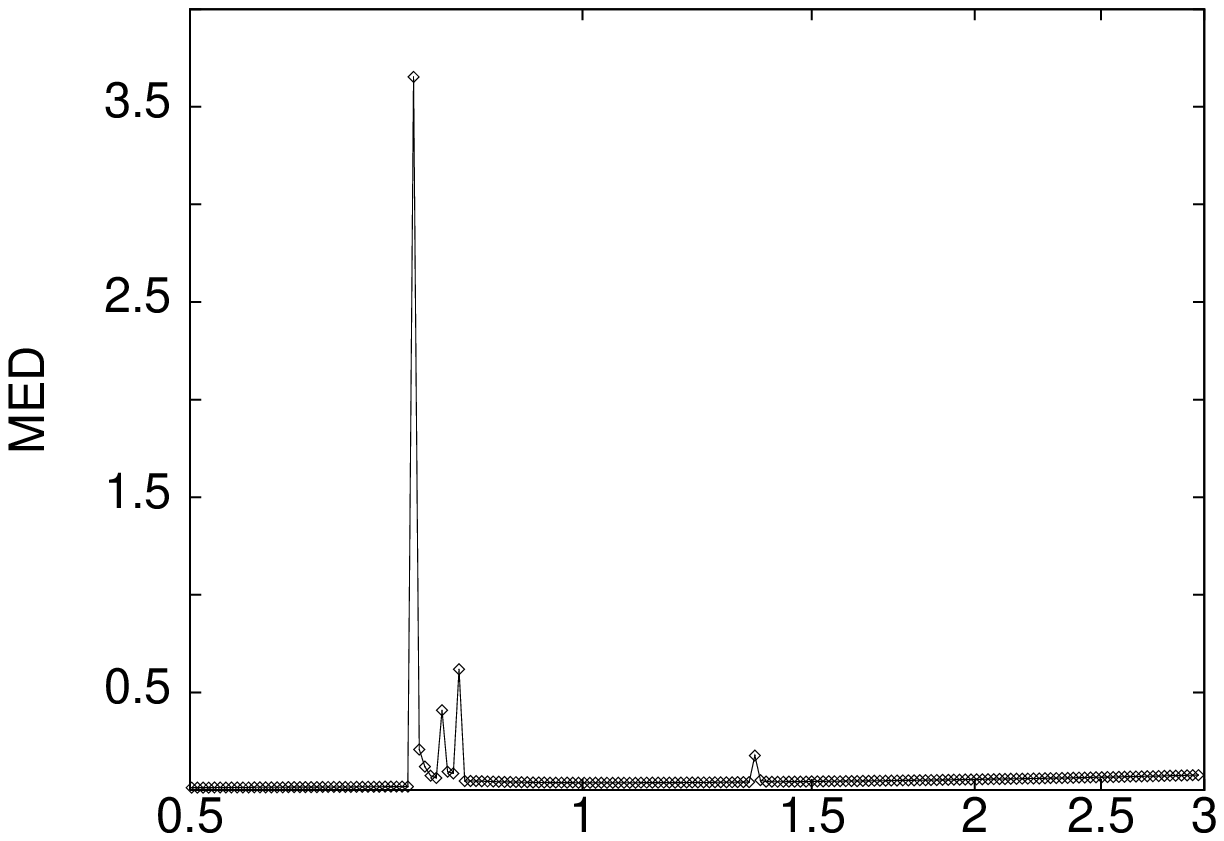}
\includegraphics[width=48mm]{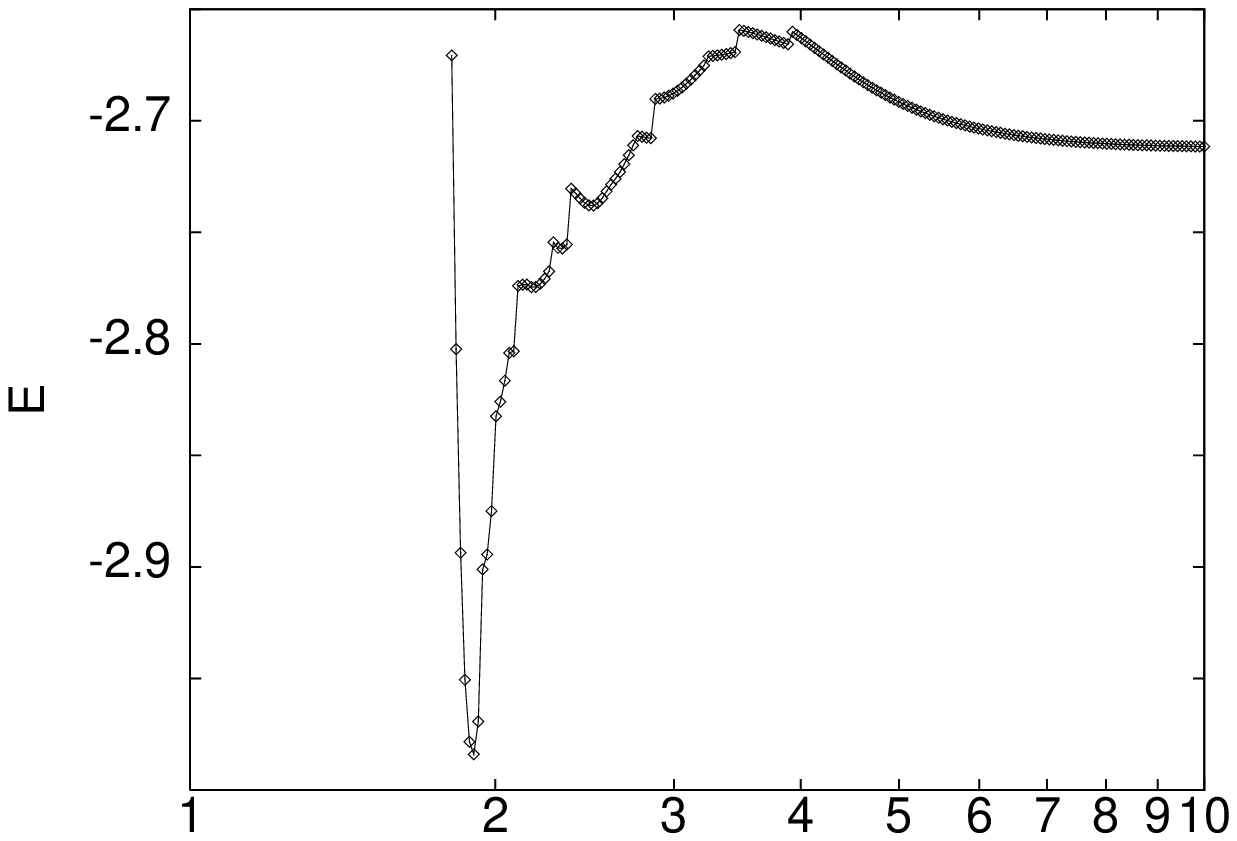}
\includegraphics[width=48mm]{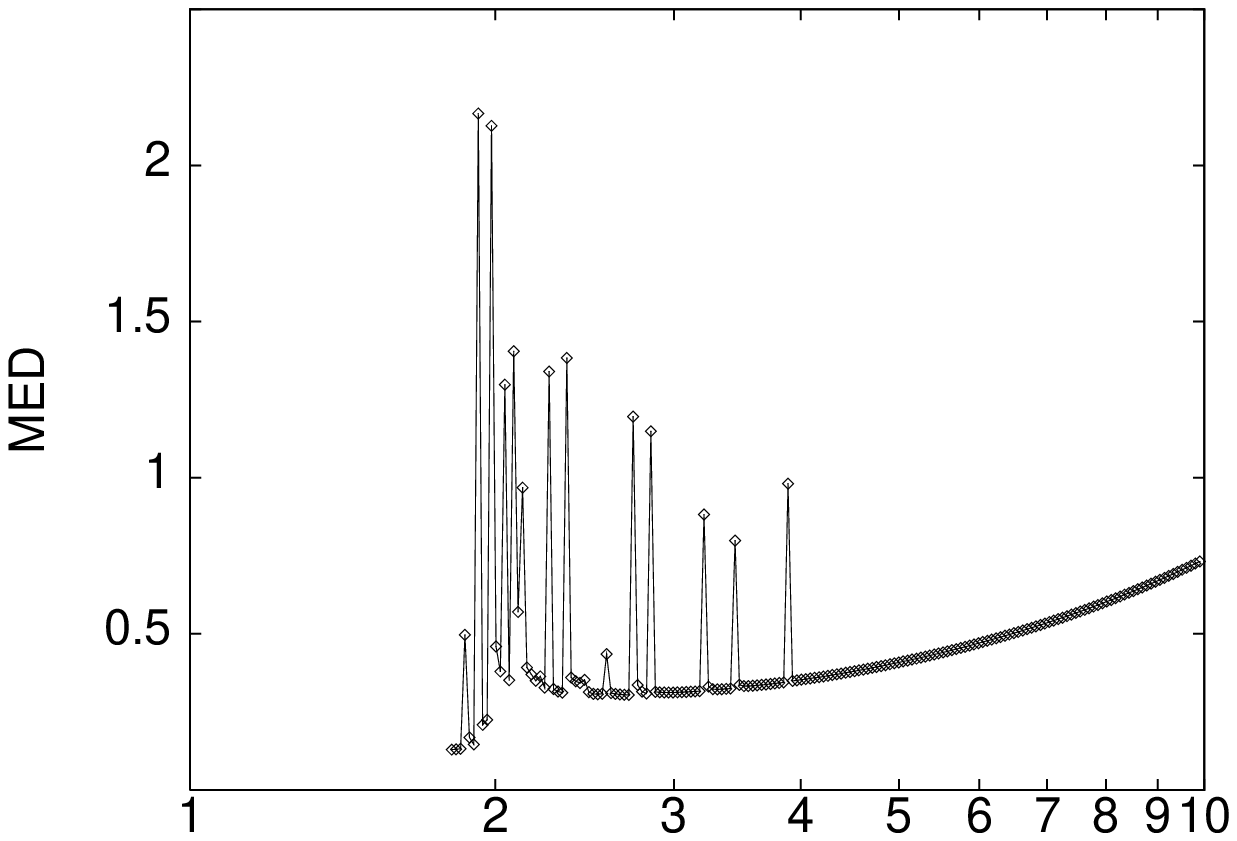}
\includegraphics[width=48mm]{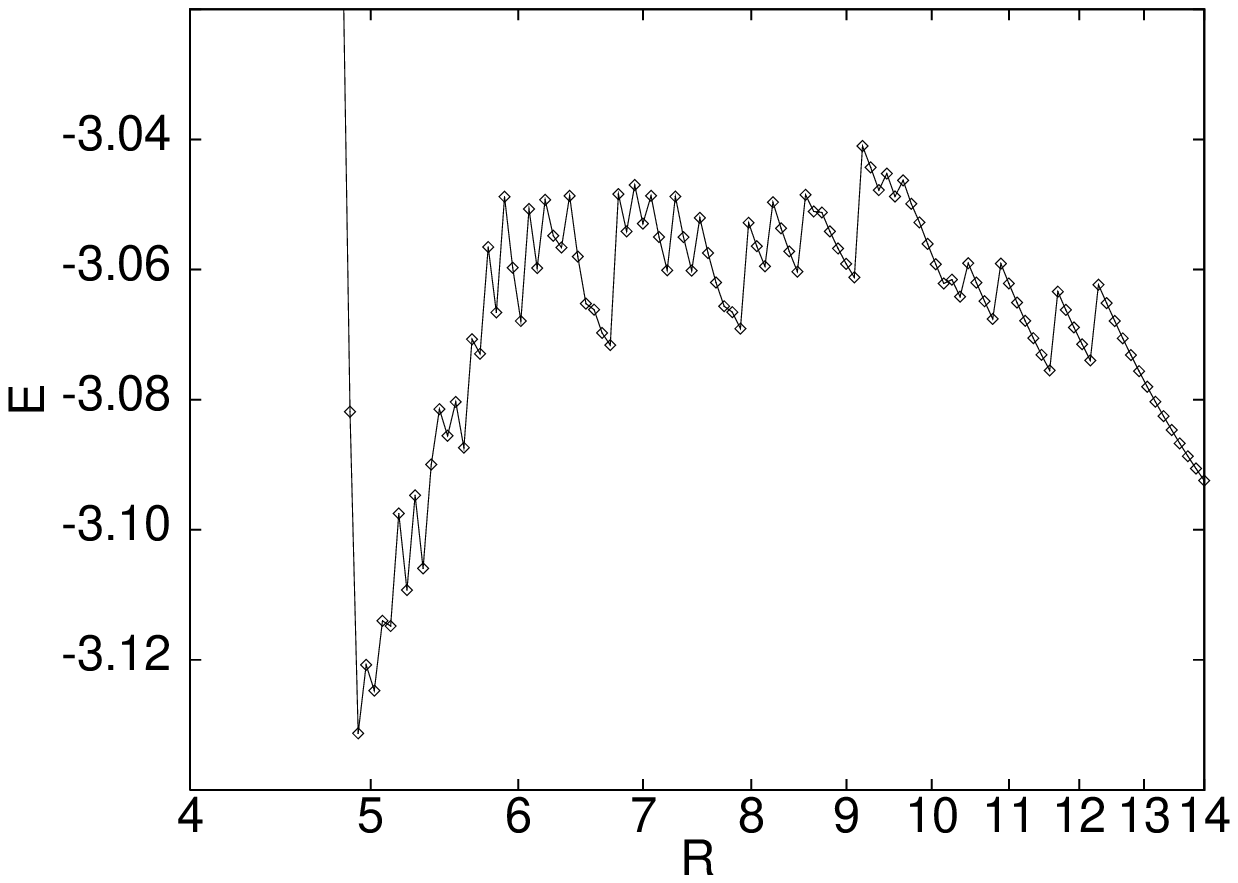}
\includegraphics[width=48mm]{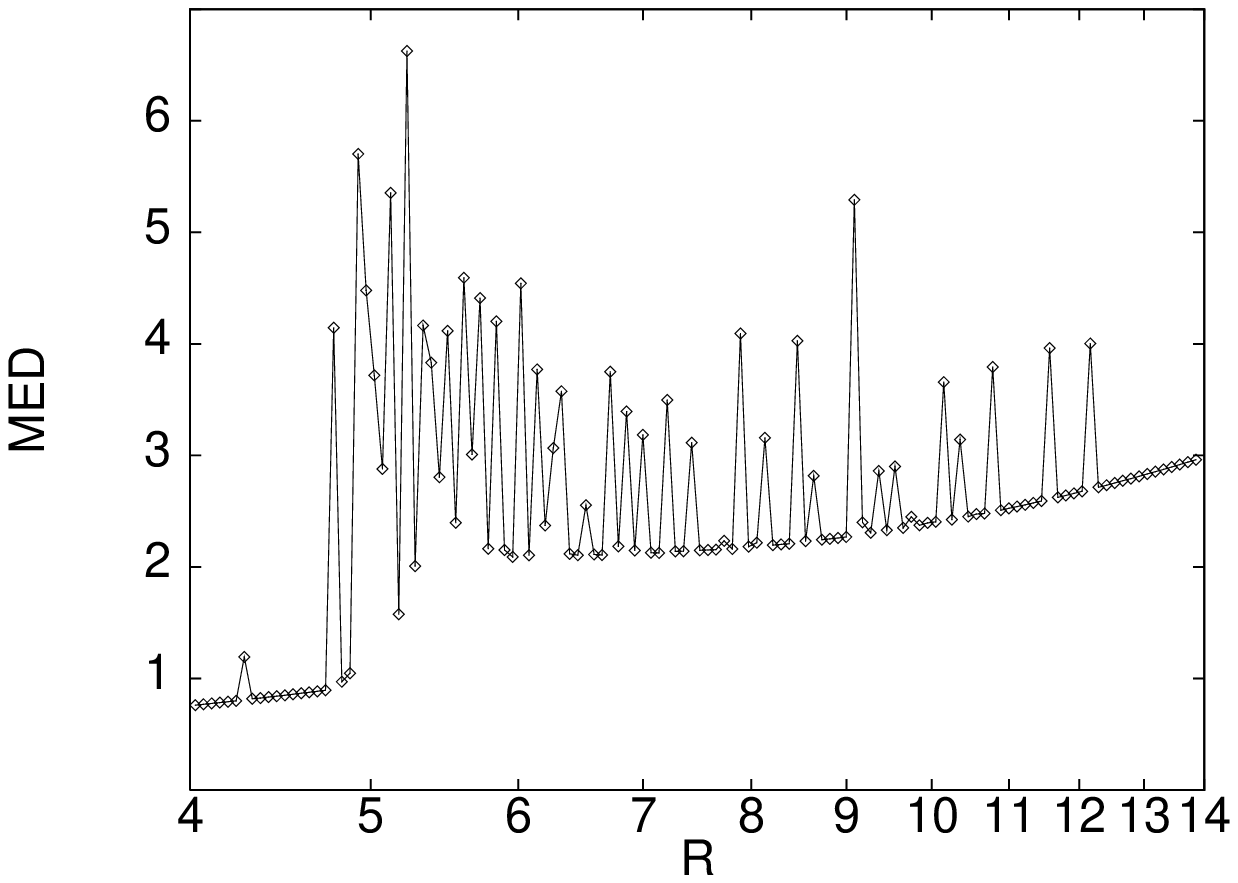}
\vspace{-7mm}
\end{center}
\caption{\sf \small \label{fig1} Data from the first series of experiments:
from top to bottom for $N=6$, 50 and 350 the mean potential energy
($E$) (at left), and the mean Euclidean distance ($MED, \hat{r}$)
traveled in configuration space during equilibration after 1\%
downward steps in curvature radius ($R$). The system was initiated in
a flat regular configuration.}
\vspace{-2mm}
\end{figure}

\noindent{\center \em Energy trends\\}  
In the first study a general energy trend is found for all $N$-values, as illustrated by three typical examples: for $N$=6, 50 and 350 in Fig.\,\ref{fig1} (left column). As expected, the most prominent feature is a deep {\em global} minimum along the followed $E-R$ path. Having started from a flat regular configuration while systematically decreasing $R$, the particles cover in close packing an increasing part of the sphere until at the closure radius $R_c$ the lowest $E$ along the $E$ - $R$ path is reached. At this stage any uncovered area -- and thus any edge -- has disappeared. We note that except for details, the structure is not biased significantly by the initial configuration.

The energy difference, $\Delta E_c$, between the $R \rightarrow \infty$ asymptotic value and the deep minimum represents the energy gain of closure. 
Additional structure is visible: secondary, local, minima and sharp downward jumps at $R > R_c$. 

Toward smaller radii, $R < R_c$, the system comes under external pressure, and the density increases. The LJ repulsive core between neighboring particles then increasingly dominates the energy, the energy rises steeply, independent of the detailed configuration. 

\begin{figure}
\begin{center}
\includegraphics[width=90mm,height=32mm]{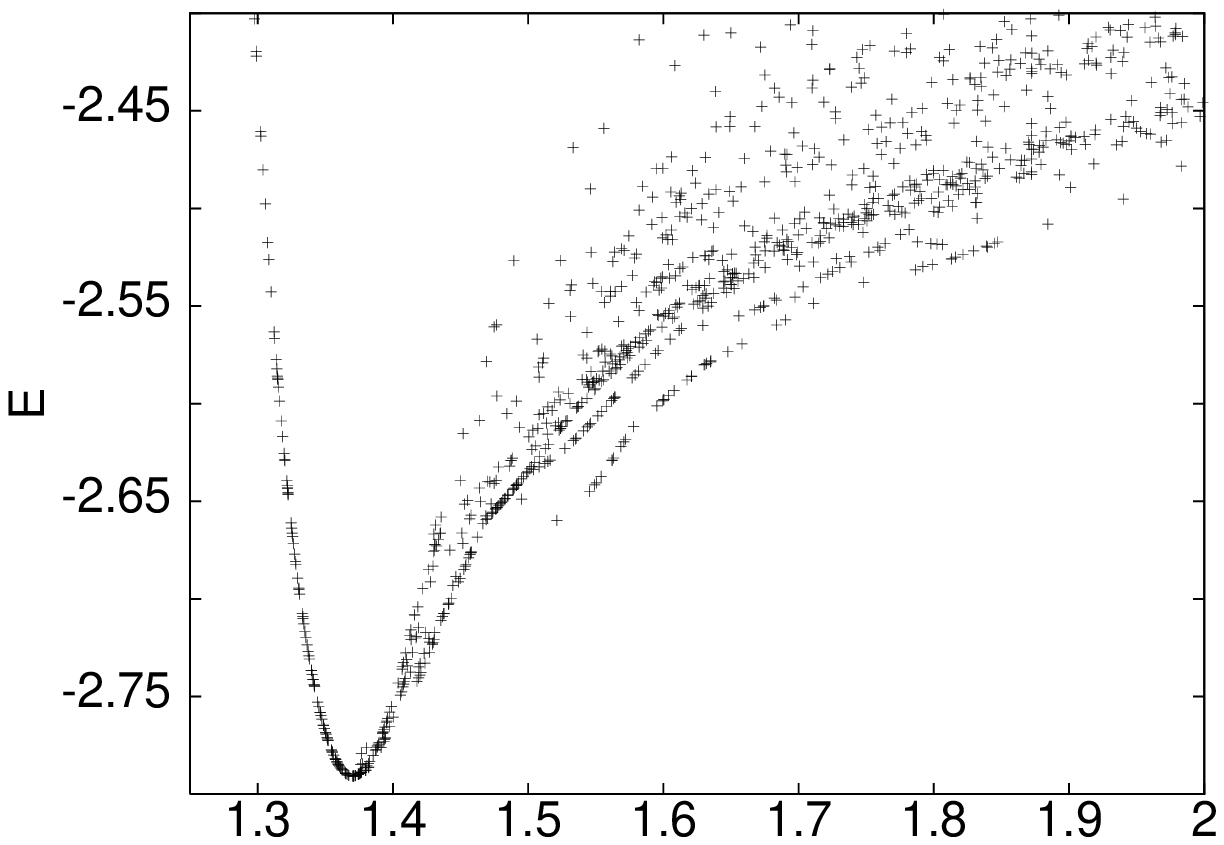}
\includegraphics[width=90mm,height=32mm]{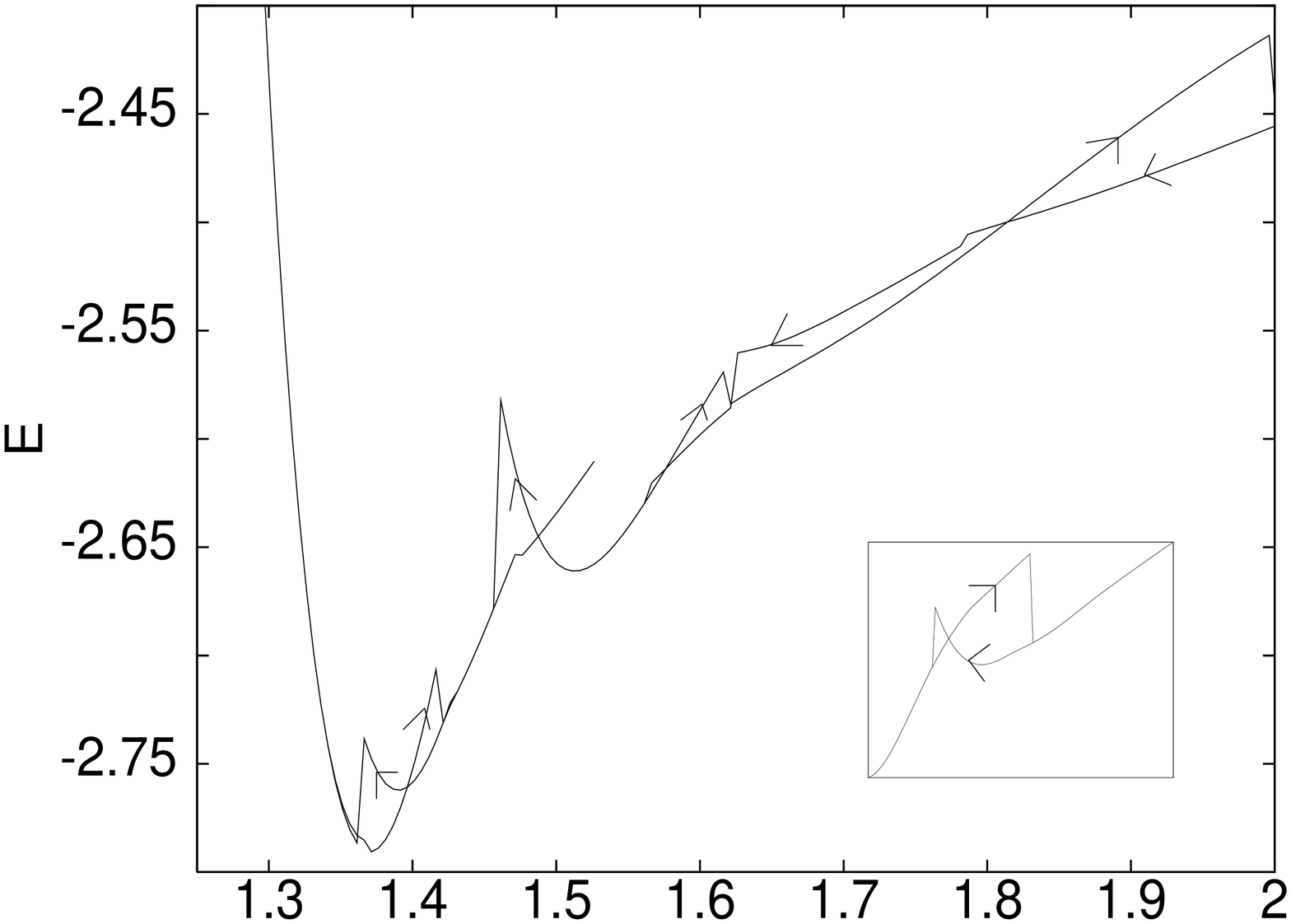}
\includegraphics[width=90mm,height=32mm]{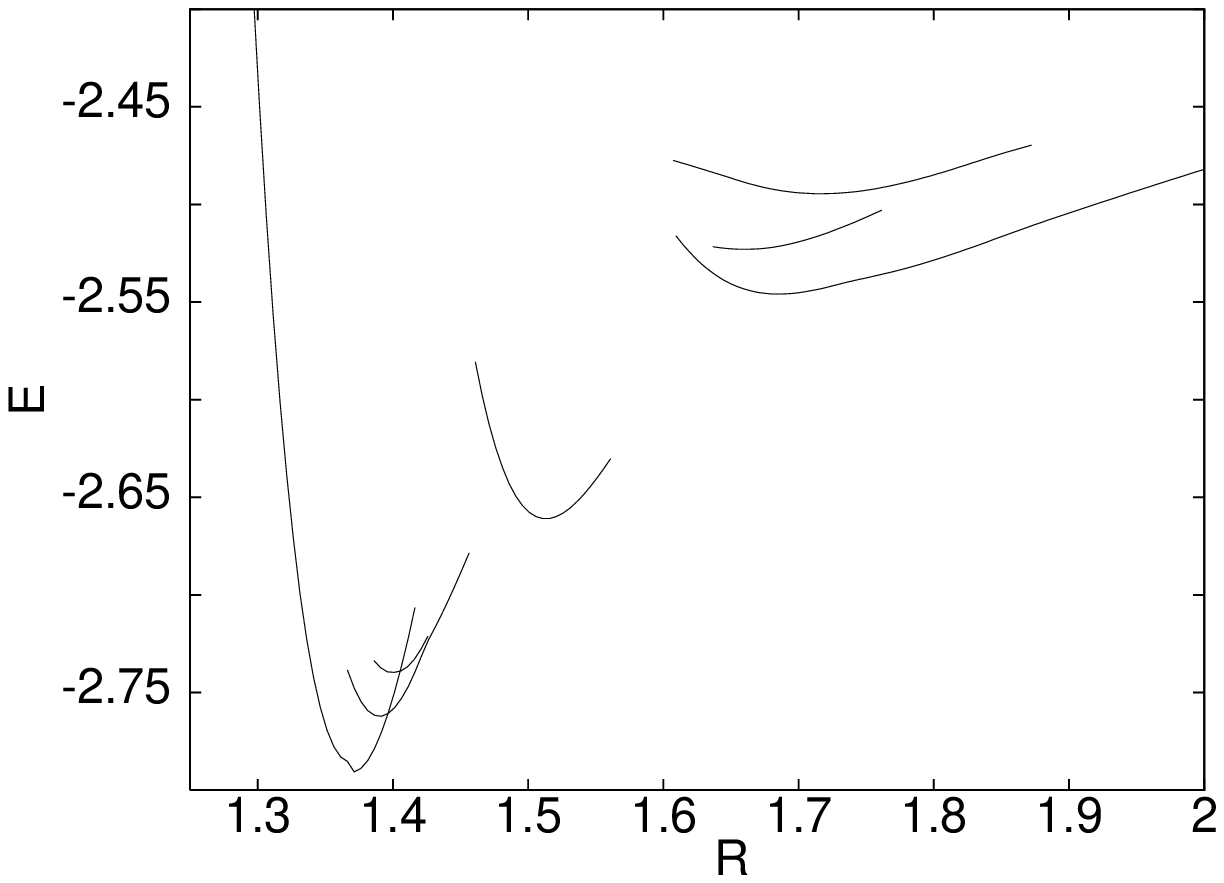}
\end{center} 
\vspace{-6mm}
\caption {\sf \small \label{fig2} Mean energy ($E$)
versus curvature radius ($R$) for $N=25$. From top to bottom: a) Second
experiment: energy minima at random fixed values of $R$, found in
unbiased searches, each
with random initial configuration, applying SA and SD; b) Third
experiment: example of tracing through $E-R$ space by 1\% steps in
$R$ using SD. Arrows indicate the $R$ step direction. The inset
shows a clear-cut case of hysteresis; c) Parts of trajectories,
exhibiting local minima as a function of $R$.}
\end{figure}

The energy of a configuration can be approximated as $E=(N_{\rm bulk}\epsilon_{\rm bulk}+N_{\rm edge}\epsilon_{\rm edge})/N$, where $N_{\rm bulk}+N_{\rm edge}=N$, $\epsilon_{\rm bulk}=-3.382$ being the lowest possible value corresponding to an infinite flat regular hexagonal lattice while $\epsilon_{\rm edge}$ is typically half that value. Due to the impossibility of fully regular hexagonal packing $\epsilon_{\rm bulk}$ is higher on a curved surface. The unfavorable edge energy forms the main driving force for curvature and shell closure, although - due to topological rearrangement - normally a barrier between flat and curved states stands in the way. Consequently, for the larger $N$-values, starting from the flat GEM configuration, indeed $dE/dR$ is negative, meaning that at sufficiently low $T$ such a flat LJ-system is locally stable against curvature. An exception is the extreme case, $N=6$, where the initially flat LJ-system can immediately gain energy by curving. A tiny jump in $E$ (at $R \approx 1.36$) goes with a discontinuity in the $dE/dR$ slope. Here already we observe a secondary minimum along the standard $E-R$ path. For $N=50$ below $R=4$, local energy minima and jumps are visible. At $R_c \approx 1.87$ the system closes over the sphere. For $N=350$ below $R=13$, many minima and many jumps show-up. At $R_c \approx 5$ the system closes. In the second experiment ($N=25$), in Fig.\,\ref{fig2}a, the independently obtained $E$-values for fixed, randomly chosen, $R$-values align along smooth lines in the $E-R$ plane.
\vspace{-4mm}
\noindent{\center \em Euclidean distance\\}
Energy jumps (Fig.1 at left) go with structural transitions visible as spikes in the mean Euclidean distance per particle per percent, $\hat{r}(R)$ (Fig.1 at right). At $R < R_c$, the particles hardly move anymore over the sphere, all flexibility being lost due to the external pressure counterbalanced by the repulsive core. The smooth background in $\hat{r}(R)$ reflects small adjustments of an essentially stable configuration after a curvature step. It can be fully suppressed by reducing the step size, which turns out to leave the transitions essentially unaffected; the sharpness of the transitions remains within a step size of even an order of magnitude smaller than applied in the current data. 
\newpage
\noindent{\center \em Closure radius\\} The radius $R_c$ at which the system closes can be predicted by requiring the packing density on the spherical surface to be (almost) the same as for a flat lattice, $R_c(N)=\sqrt{\frac{N}{4\pi}\cos{(\frac{\pi}{6})}}$.
For $N$=6, 25, 50 and 350 it follows that $R_c(N)$=0.64, 1.3, 1.9 en 4.9 respectively. These values agree with the data (Fig.\,1 left) except for small $N$ where the packing is far less optimal than for a flat GEM lattice.
\vspace{-1mm}
\noindent{\center \em Structural transitions and topological defects\\} Energy jumps between different trajectories involve a major global rearrangement. 

With successively applied small steps in curvature followed by
relaxation, the system usually undergoes only minor local redistributions, while strain accumulates in parts of the lattice. At some curvature steps, however, an avalanche of sequential displacements over a major part of the system is triggered by
local instabilities, while releasing much of the built-up strain. A small change in $R$ may thus enforce a structural transition to a configuration, in which both the strain and the
2D-topological (defect) structure have been drastically altered. An extensive systematic study of the topological defect structure as a function of $N$ up to 200 is available\,\cite{voogd-thesis}.
 
The transitions come along with incorporation or removal of defects as a function of
curvature. Energy barriers are present between configurations with different topological structure. The transition to take place requires the system to climb the barrier to a threshold for starting an avalanche of particle moves. The threshold $R$-value depends on the barrier side, causing hysteresis. 

Starting from an essentially circular patch with flat regular packing, the `unstable' radius $R_u$ where - with increasing curvature - the first structural transition occurs, is modeled 
as $S * 2n_{\rm max}\arcsin(\alpha) = \sin(2n_{\rm max} \arcsin(\alpha))$ \cite{voogd-thesis}, where $\alpha = 1/(2R_{\rm u})$, $n_{\rm max}$ is essentially the largest completed hexagonal ring, and where S can be taken from the data for a single specific $N$-value 
(see Table\,\ref{table1}). 
\vspace{-4mm}
\noindent{\center \em Open configurations\\} The points in Fig.\,\ref{fig2}a at large $R$ cover a broad range in energy, corresponding to a great variety of open configurations and
edge-arrangements. As $R$ decreases, the edge becomes smaller, and the variation decreases. Any flexibility essentially disappears below $R_c$, due to the strong constraints for a closed spherical configuration.

\begin{table}[t]
\begin{center}
\begin{tabular}{rccc}
 $N$ & $n~(max)$ & $R_{u}~(data)$ & $R_{u}~(model)$ \\
\hline
6   &  1   &    1.37  &  1.34 \\
50  &  3   & 3.95 & 4.11 \\
350 &  9   &    12.3  &  12.3 \\
500 &  11  &    15.6  &  15.0 \\
\end{tabular}
\end{center}
\vspace{-4mm}
\caption{\sf \small \label{table1} The radius $R_{\rm u}$ where
the trajectory becomes unstable for increasing curvature starting
from a flat regular distribution for $N$=6, 50, 350 and 500,
compared with estimates from the simple model with $S$ calibrated at $N=350$.}
\end{table}
\vspace{-2mm}
\noindent{\center \em Secondary minima\\} 
Fig.\,\ref{fig2}b shows $E-R$ trajectories from the third experiment, starting from a specific configuration (near the minimum at $R \approx$ 1.5) from the second experiment. Here the
system is traced up-and-down in $R$. Indeed, the trajectories in the $E-R$ plane connect unbiased solutions, and -- like in the first experiment -- jumps and secondary minima occur.
The secondary minima visible in Fig.2c, are obtained starting from   configurations of the second experiment (Fig.2c), and tracing in $R$ up and down  until a jump occurs.  The minima have a typical depth of 10\% of the closure energy $\Delta E_c$. At non-zero $T$ the thermodynamic significance of such minima should be judged with respect to both kT and $\Delta E_c$. 

\noindent{\center \em Variable $N$\\} 

{The finding of distinct locally-stable open configurations for fixed $N$ raises the expectation that any discrete curvature may remain, or change smoothly, during growth when particles are added.  A study of this type\,\cite{voogdp157} indeed indicates that discrete locally-stable configuration trajectories exist as a function of both $R$ and $N$, where the bulk packing remains essentially the same.}

\vspace{2mm}
In conclusion, for fixed-$N$ two-dimensional spherical Lennard-Jones systems at zero temperature, the global energy minimum with decreasing curvature radius $R$ is approached through sharp transitions with major rearrangements. These transitions bring-in topological defects connected with curvature. The curvature range of events relevant for self-assembly, $R_u$ - $R_c$ is consistent with simple models. Apart from the closed (global) minimum energy configuration, secondary, local, minima show up at larger $R$-values, with an open configuration. This phenomena  -- here shown for $N=25$ -- occurs naturally as a consequence of optimal packing topologies. During growth toward a complete shell such minima can capture `threads' of steady curvature along with growth in $N$, while staying in equilibrium.
The present results and methods can help guide further generic studies of the self-organization in complex spherical molecular systems. 

\vspace{2mm}
The authors are grateful to D. Frenkel (FOM-AMOLF)
for many fruitful suggestions, and to M. Livny (UW-Madison) for indispensable support on Condor High Throughput Computing. Part of this work has been funded by the FOM and NWO science organizations in The Netherlands.

\end{document}